\documentclass{aa}
\usepackage{graphics}
\begin{document}
\thesaurus{03(09.03.1; 11.09.1M81)}
\title{Identification of molecular complexes in \object{M 81}}
\author{N. Brouillet\inst{1}
  \and M. Kaufman \inst{2}
  \and F. Combes \inst{3}
  \and A. Baudry \inst{1}
  \and F. Bash\inst{4}}
\institute{
 Observatoire de l'Universit{\'e} Bordeaux 1, CNRS URA n$^{\circ}$ 352, B.P.
 89, F-33270 Floirac, France
\and
 Department of Physics, The Ohio State University, 174 West 18th Avenue,
 Columbus, OH 43210, USA
\and
 Observatoire de Paris, 61 avenue de l'Observatoire, 75014 Paris, France
\and
 Department of Astronomy, University of Texas at Austin, Austin, TX 78712, USA}
\offprints{N.~Brouillet}
\date{Received August 6/ Accepted December 22, 1997}
\maketitle
\begin{abstract}

We report about high spatial resolution observations made with the IRAM Plateau 
de Bure interferometer of the $^{12}$CO($J = 1\rightarrow 0$) emission from a 
$1.1\,\times\,1.1$\,kpc plane-of-sky field on a spiral arm of Messier 81.  With 
a beam of $5''\approx$\,90 pc, we identify 6 giant molecular cloud complexes 
with virial masses of $\approx 10^6$\,M$_\odot$, including one associated with 
a giant HII region.  The deduced $N\,({\rm H}_2$)/$I_{\rm CO}$ ratios are about 
3 times larger on average than those measured near the solar neighborhood, 
suggesting that the complexes are not self-gravitationally bound except, 
possibly, for the complex associated with the giant HII region; they could be 
the average of several clouds of mass a few 10$^5$\,M$_\odot$ and diameter $\le 
$100\,pc.  The linewidths are very narrow with respect to the measured sizes, 
so that the size-linewidth relation for \object{M 81} clouds is very different 
from that in the Milky Way.  The narrow linewidths imply smaller virial masses 
than for Galactic complexes of the same size, and this is consistent with the 
weaker CO emission from the GMCs in \object{M 81}.  The low velocity dispersion 
suggests a lower mean volume density in the cloud and, possibly, a smaller 
scale height of the molecular gas than in Galactic clouds of the same size.  
Comparison of the interferometer and single-dish line profiles indicates that, 
at most, 30\% of
the single-dish emission in this field is from a widespread distribution of 
small clouds, and thus the population of molecular clouds is rather different 
from that in the Milky Way.

The H$_2$ surface density in \object{M 81} is low: although the region studied 
here is one of the richer molecular regions in the disk, the molecular surface 
density is much smaller than the interarm regions of \object{M 51} for example.  
The HI gas dominates and can explain most of the extinction seen at optical 
wavelengths in this field.  In some other fields, the HI gas cannot explain the 
observed extinction, but previous lower resolution observations detected little 
or no CO there.  The present high resolution observations imply that the 
molecular medium in \object{M 81} differs from that in the Milky Way.


\end{abstract}

\section{Introduction}

To understand the star formation activity at both large and small scales, one 
must know the distribution of the atomic and molecular gas on both scales.  The 
importance of molecular clouds as star cradles and the easier study of rich 
molecular galaxies like \object{M 51}, made low CO emitting galaxies, like 
\object{M 81}, puzzling -- as long as one relies on CO as a molecular gas mass 
indicator.  However Sauty et al.  (\cite{sauty}) recently showed that a typical 
Sa to Sc spiral galaxy has three to four times less molecular gas than atomic 
hydrogen.  This result is different from the study of Young \& Knezek 
(\cite{young}) who find a higher H$_2$/{\hbox{HI}} ratio in earlier type 
galaxies compared to late types.  The discrepancy may be due to the selection's 
criteria as the galaxies in the sample of Young \& Knezek are mostly FIR and 
optically bright objects and thus more probably CO-bright objects.

In \object{M 81} the $^{12}$CO($J = 1\rightarrow 0$) emission has been mapped 
from the nucleus out to a radius of 7$\,.\!\!^{\prime}$5 and the inferred mass 
of molecular hydrogen is $\sim 10^8$\,M$_\odot$ (Brouillet et al.  
\cite{brouillet2}; Sage and Westpfahl \cite{sage}), assuming a standard 
$N\,({\rm H}_2$)/$I_{\rm CO}$ ratio.  The HI mass within the same radius is 
5\,$10^8$\,M$_\odot$ (Hine \cite{hine}) which leads to a molecular to atomic 
gas ratio of 0.2, a bit less than the mean value (0.3) derived by Sauty et al.  
for Sab galaxies, and much less than the average value (2) derived by Young \& 
Knezek.

However even if \object{M 81} does not appear anymore so atypical, the 
distribution of molecular gas with respect to the other tracers of star 
formation still raises questions.

Reichen et al.  (\cite{reichen}) from the distribution of UV 2000\,\AA\ 
emission find that massive star formation apparently does not occur 
preferentially at the locations of the highest gas surface density.  
Furthermore in our CO surveys (Brouillet et al.  \cite{brouillet1},  
\cite{brouillet2}) we found a peculiar lack of emission near the most luminous 
\hbox{HII} regions, which would confirm that they are efficient in dissociating 
the molecular gas or in pushing it out (Kaufman et al.  \cite{kaufmana}).  But 
there are still a few \hbox{HII} regions where the visual extinction is too 
high if only the atomic gas is taken into account and which must then have 
associated molecular gas.  The dilution factor in the IRAM 30\,m beam can 
explain the observed low $N\,({\rm H}_2)$ which would mean that the CO is 
highly concentrated on these \hbox{HII} regions.

Other evidence that the CO measurements may underestimate the molecular gas 
mass is provided by the dust lanes.  Kaufman et al.  (\cite{kaufmanb}) deduce, 
from measurements of the visual extinction and the {\hbox{HI}} column density, 
that the dust filaments in \object{M 81} are composed mainly of molecular gas.  
Though CO emission appears to be associated with the dust regions on a large 
scale, we did not detect any strong emission with the 30\,m towards two 
particular dust lane positions of high optical depth (A$_v$ $\ge$\,16) 
(Brouillet et al.  \cite{brouillet2}).  The mass of molecular gas is at least a 
factor of 3 to 5 times lower than that inferred by Kaufman et al.  
(\cite{kaufmanb}).  This inconsistency suggests that: (i) the gas-to-dust ratio 
is different from the Galactic value -- but it seems unlikely that it is as low 
as required to explain the discrepancy; or (ii) the $N\,({\rm H}_2$)/$I_{\rm 
CO}$ ratio is different from the Galactic value; or (iii) the molecular clouds 
have a low excitation temperature $T_\mathrm{ex}$.

Indeed Allen and Lequeux (\cite{allen}) have detected extended molecular clouds 
associated with dark dust clouds in the inner disk of \object{M 31}.  These 
molecular clouds have a very low excitation temperature and display much 
fainter CO emission than Galactic GMC's.  In such a case the CO line intensity 
cannot so straightforwardly be used as an indicator of the molecular mass.

High spatial resolution observations are needed to precisely locate the 
molecular clouds in \object{M 81} with respect to the dust filaments and 
\hbox{HII} regions.  Furthermore as the CO emission is weak, we need the most 
sensitive interferometer.

 From our previous 30\,m observations we identified the best field in order to
detect molecular complexes in \object{M 81}. In the N\,5 field (Brouillet et al.
\cite{brouillet2}) CO emission follows the East-West dust lane and at the
($-20^{\prime\prime},-10^{\prime\prime}$) position the CO profile is double
peaked. The \hbox{HII} region 138 (Kaufman et al. \cite{kaufman}) lies
30$^{\prime\prime}$ farther to the West. This HII region is not representative
of the HII regions with too much extinction; half of the observed HI in its
direction is already sufficient to account for its extinction. We observed a
position on the dust lane, covering the
($-20^{\prime\prime},-10^{\prime\prime}$) 30\,m position and most of \hbox{HII}
region 138.

In this article we present the $^{12}$CO($J = 1\rightarrow 0$) observations 
(Sect.~\ref{results}) and then discuss the inferred $N\,({\rm H}_2$)/$I_{\rm 
CO}$ ratio in \object{M 81} (Sect.~\ref{X}).  In Sect.~\ref{HII} we comment on 
the association of the molecular complexes and the HII region.

\section{Observations} \label{observations}

CO\,($J$=1$-$0) line observations at 115 GHz were carried out towards \object{M 
81} with the IRAM Plateau de Bure Interferometer in August and September 1995.  
We used the compact CD configuration which led to a synthesized beam of 
5$\,.\!\!^{\prime\prime}$1 $\times$ 4$\,.\!\!^{\prime\prime}$8; this 
corresponds to 89\,pc $\times$ 84\,pc at the adopted distance of 3.6\,Mpc 
(Freedman et al.  \cite{freedman}).  The primary beam is $44^{\prime\prime}$ 
HPBW. The four antennas were equiped with SIS receivers and system temperatures 
ranged from 300 to 700\,K. 1044+719 and 0954+658 were used as phase and 
amplitude calibrators.  3C273, 3C84, NRAO530 and 3C454.3 were used as bandpass 
and flux calibrators.  The six units of the correlator were split as follows: 
two units covering 160\,MHz each were used for the continuum, two 256 channels 
units and two 128 channels units provided a velocity coverage of 
155\,km\,s$^{-1}$ and 258\,km\,s$^{-1}$ respectively, centered at 
$V_\mathrm{lsr} = 171$\,km\,s$^{-1}$ ($V_\mathrm{lsr} 
=V_\mathrm{helio}+6$\,km\,s$^{-1}$).  The channel separation was 0.156\,MHz and 
0.625\,MHz respectively thus giving an effective velocity resolution of 
0.65\,km\,s$^{-1}$ and 2.6\,km\,s$^{-1}$ (since the channels are not 
independent).

The phase center of the observations was $\alpha_{2000} = 9^{\rm h} 55^{\rm m}
00\,.\!\!^{\rm s}6$, $\delta_{2000} = 69\degr 08^{\prime}
44\,.\!\!^{\prime\prime}2$.  The coordinates were afterwards transformed into
the B1950 system and the phase center in this system is then $\alpha_{1950} =
9^{\rm h} 50^{\rm m} 54\,.\!\!^{\rm s}0, \delta_{1950} = 69\degr 22^{\prime}
56\,.\!\!^{\prime\prime}0$.

The total on-source integration time was about 34 hours.  No continuum emission 
was detected at a 5\,mJy level (3$\sigma$).  The line maps were cleaned using 
the CLARK algorithm and then corrected for primary beam attenuation.  We 
achieved a sensitivity of $\sim$0.5\,K\,km\,s$^{-1}$ (1$\sigma$) at the center 
of the line integrated-intensity map (map integrated in velocity over a 
52\,km\,s$^{-1}$ range) and $\sim$2.5\,K\,km\,s$^{-1}$ (1$\sigma$) on the 
outside border.  The flux has been converted from Jy/beam to K by multiplying 
by 3.8\,K/Jy.

We also made CO\,($J$=1$-$0) observations with the IRAM 30\,m radiotelescope in 
November 1996.  Due to the minimum spacing of the Plateau de Bure antennas 
(35\,m), structures more extended than 15$^{\prime\prime}$ cannot be detected.  
The 30\,m telescope can thus perfectly provide the short spacings data to 
complete the interferometer set.  We used two SIS receivers in parallel; the 
system temperature was about 500\,K. Two 1\,MHz filterbanks and an 
autocorrelator gave velocity resolutions of 2.6 and 1.3\,km\,s$^{-1}$ 
respectively.  The half power beamwidth was $23^{\prime\prime}$.  Unfortunately 
due to bad weather we could not complete our single-dish map in order to add it 
to the interferometer data.  However, we obtained a map of size 
20$^{\prime\prime}$\,$\times$\,20$^{\prime\prime}$ with 8$^{\prime\prime}$ 
spacing and a sensitivity of 3$\sigma\sim$80\,mK (at the center) and 110\,mK 
(on the edge) for 2.6\,km\,s$^{-1}$ resolution.  The pointing error 
is estimated to be 5$^{\prime\prime}$ at most.

\section{Results} \label{results}

\begin{figure}
\includegraphics{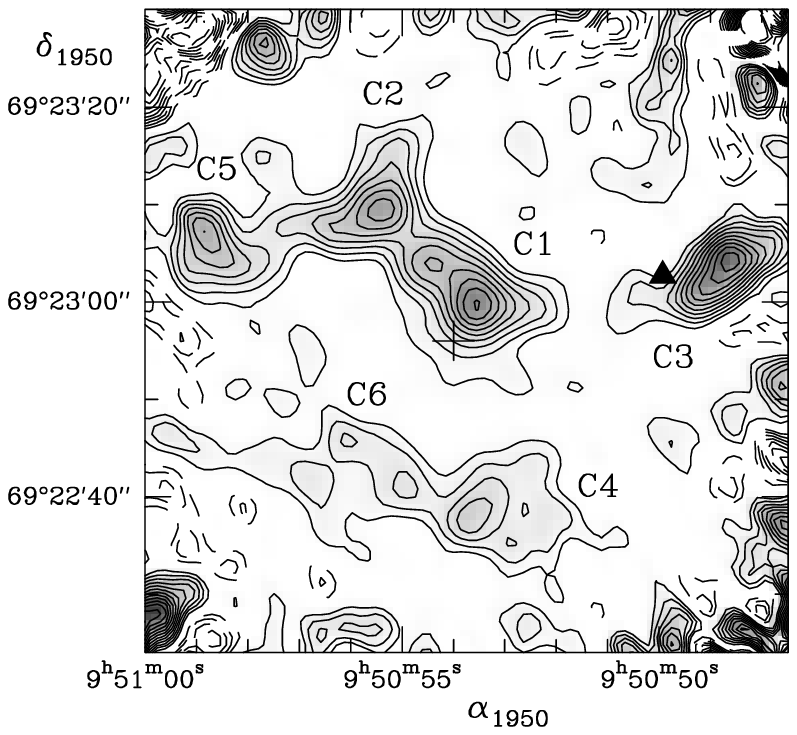}
\caption[]{CO(1-0) line integrated-intensity map after
correction for primary-beam attenuation;
the map is integrated in velocity
from 145 to 197\,km\,s$^{-1}$.  The contour spacing is
1\,K\,km\,s$^{-1}$ (the zero level is suppressed), corresponding to 2 times
the noise level at the center of the map, and the clean beam is
5$\,.\!\!^{\prime\prime}$1 $\times$ 4$\,.\!\!^{\prime\prime}$8.  The cross
refers to the phase center of the observations at $\alpha_{1950} = 9^{\rm h}
50^{\rm m} 54\,.\!\!^{\rm s}0, \delta_{1950} = 69\degr 22^{\prime}
56\,.\!\!^{\prime\prime}0$.  The filled triangle marks the center of the HII
region 138.}
\label{integrated-map}
\end{figure}

\begin{figure*}
\includegraphics{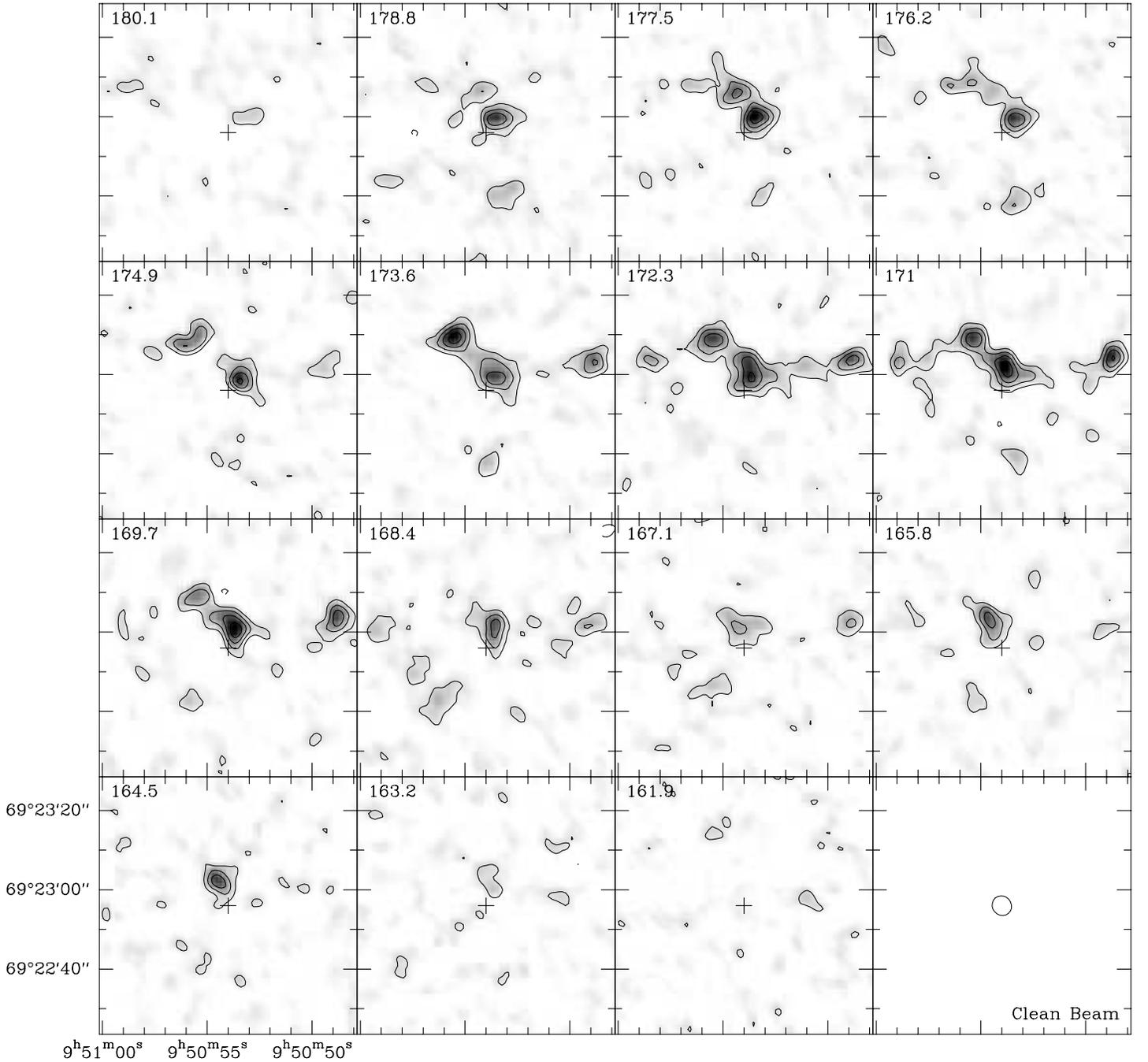}
\caption[]{Channel maps of the CO(1-0) line emission before
correction for primary-beam attenuation.  The contour spacing is 40\,mJy/beam
or 0.15\,K (the zero level is suppressed) and the clean beam is
5$\,.\!\!^{\prime\prime}$1 $\times$ 4$\,.\!\!^{\prime\prime}$8.  The cross
refers to the phase center of the observations.  The lsr center velocity of
each map is written in the upper left corner and the velocity spacing is
1.3\,km\,s$^{-1}$.  The rms noise before correction for primary-beam
attenuation is 55\,mK.}
\label{channel}
\end{figure*}

Figure~\ref{integrated-map} displays the line integrated intensity map and
Figure~\ref{channel} the channel maps.  The field, which is located in the
northern spiral arm, displays a complex molecular structure and there is no
clear velocity structure.

 From fig.~\ref{channel} we identified 6 molecular complexes, noted C1 to C6 on 
fig.~\ref{integrated-map} and in Table~\ref{table}.  They were defined as 
structures appearing on at least 2--3 adjacent channels.  These complexes range 
in integrated CO flux from 1.3 Jy\,km\,s$^{-1}$ for C6 to 5.0 Jy\,km\,s$^{-1}$ 
in the case of C1 and C3, and in effective radius from 78 pc for C4 to 105 pc 
for C1.  Complex C1 is located near the phase center, but C3 and C5 lie 
well beyond the half power point of the primary beam.

\begin{figure}
\includegraphics{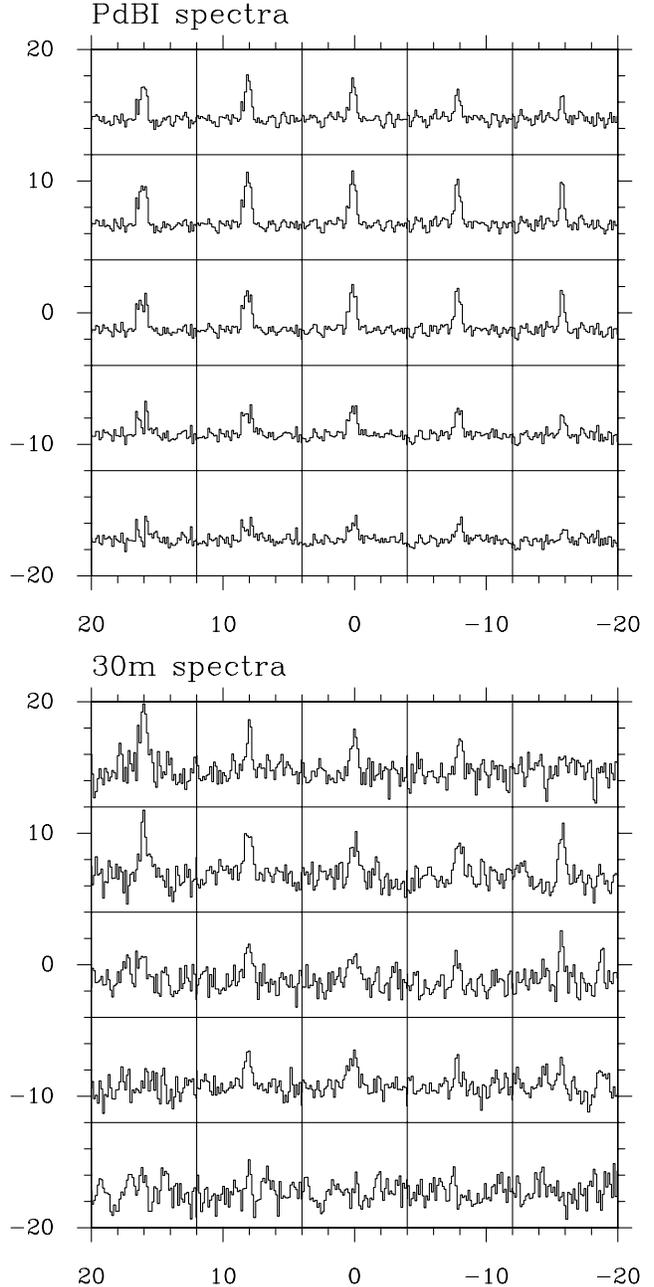}
\caption[]{Interferometric (top) and single-dish (bottom) maps convolved to 
the same spatial resolution (23$^{\prime\prime}$).  The (0,0) position is the 
phase center of the interferometer data; the offsets are in arcsec.  The 
velocity scale is 100 to 250\,km\,s$^{-1}$ and the intensity $T_{\rm mb}$ scale 
is -0.1 to 0.2\,K.}
\label{pdbi-30m}
\end{figure}

In order to estimate the extended emission missed by the interferometer, we 
compared the 30\,m spectra to the interferometric observations convolved to the 
same spatial resolution (see fig.  ~\ref{pdbi-30m}).  The two data sets agree 
very well in velocity and line intensity except for the 
(16$^{\prime\prime}$,16$^{\prime\prime}$) position.  As this position is at the 
border of the interferometric map and because of the noise, we cannot conclude 
if a GMC within the 23$^{\prime\prime}$ beam of the 30\,m was too far from the 
phase center to be detected by the interferometer or if extended emission is 
missing from the interferometer map at this peculiar position.  Over the entire 
observed field the difference of flux is in the noise of the spectra and the 
missing flux can be estimated to 0--30\% because of the calibration
uncertainties.  Thus there is no strong structure more extended than 
15$^{\prime\prime}$ (300\,pc) and no widespread distribution of small molecular 
clouds.

\begin{figure}[ht]
\includegraphics{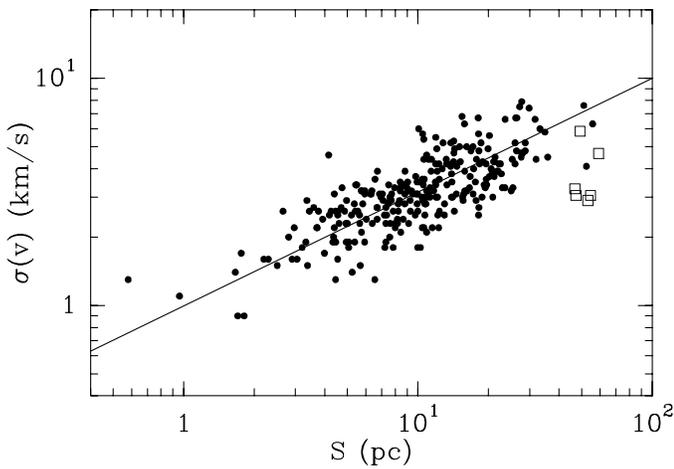} 
\caption[]{Molecular cloud velocity dispersion $\sigma (v)$ as a function of 
size $S = 0.52 R_{\rm e}$ as defined in Solomon et al.  (\cite{solomon}), see 
their figure 1.  The dots represent the Galactic clouds and the corresponding 
fitted line is $\sigma (v)=S^{0.5}$\,km\,s$^{-1}$\,pc$^{-1}$.  The open squares 
represent the \object{M 81} complexes.  For the \object{M 81} complexes the 
uncertainty in size $S$ is $\pm 9$\,pc, and the uncertainty in the velocity 
dispersion is $\pm $1\,km\,s$^{-1}$.}
\label{dv-size}
\end{figure}

We note that the linewidths of the complexes are narrow compared to Galactic 
complexes of the same size.  The \object{M 81} linewidths plotted on the 
velocity dispersion versus size diagram of Solomon et al.  (\cite{solomon}) are 
well below those for the Galaxy (fig.~\ref{dv-size}).  For the structures found 
by the interferometer, $R \sim$100\,pc, we should expect a width of 
15--20\,km\,s$^{-1}$ typically, while we find 7--14\,km\,s$^{-1}$.  The 
linewidths of the clouds in this paper (mostly 7\,km\,s$^{-1}$ for a $5''$ 
beam) scale up to the linewidths observed for \object{M 81} clouds by Brouillet 
et al.  (\cite{brouillet2}) with the 30\,m $23''$ (400pc) and $12''$ beams (of 
the order of 5 to 20\,km\,s$^{-1}$) via, approximately, the classical relation 
$\Delta V \propto$ size$^{1/2}$.  This assumes the linewidths are due only to 
self-gravitation and turbulence of the molecular clouds (not to rotational 
gradients in the beam).  In magnitude the linewidths in both cases lie below 
the Solomon et al.  (\cite{solomon}) relation (or extrapolated relation) for 
Milky Way clouds, e.g., with a $23''$ beam, we should have expected at least 
30\,km\,s$^{-1}$ for clouds similar to Galactic clouds, if the clouds filled 
the beam.  But it is uncertain to extrapolate the Solomon et al.  
(\cite{solomon}) relation to structures of 400\,pc in size, the observed 
linewidths must include the velocity gradient across the beam, and if clouds do 
not fill all the beam, they will not show the complete V-gradient.

The \object{M 81} linewidths are also narrow compared to those of the two 
molecular complexes identified by Viallefond et al.  (\cite{viallefond}) in 
\object{M 33} and to the molecular complex mapped by Casoli et al.  
(\cite{casoli}) in \object{M 31}.  On a larger scale, Table 2 of 
Garc\'{\i}a-Burillo et al.  (\cite{santiago}) lists the CO linewidth-size 
relation for external galaxies and, except for \object{NGC 628}, \object{M 81} 
appears to have narrower lines.  Although a weaker turbulence, perhaps 
resulting from a less active star formation throughout the \object{M 81} disk, 
could explain the narrow lines, the star formation activity in \object{M 81} 
(and \object{NGC 628}) is generally stronger than in \object{M 31}.  Another 
possible explanation is the CO abundance: if the clouds are less optically 
thick, the lines can be narrower.

\section{The $N\,({\rm H}_2)$/$I_{\rm CO}$ ratio} \label{X}

The value of the $X=N\,({\rm H}_2)/I_{\rm CO}$ ratio is still a subject of 
debate, especially because the ``standard'' value 
($X=2.3\,10^{20}$\,H$_2$cm$^{-2}$/(K\,km\,s$^{-1}$), Strong et al.  
\cite{strong}) adopted for the Milky Way does not seem to be valid everywhere 
in the Galaxy.  For example Digel et al.  (\cite{digel}) recently derived a 
lower value $X=1.06\,10^{20}\,$H$_2$cm$^{-2}$/(K\,km\,s$^{-1}$) towards Orion.  
The determination of $X$ in other galaxies is thus of great interest.

The resolution of molecular observations is increasing due to millimeter 
interferometers which allow us to resolve molecular complexes in nearby 
galaxies.  The $X$ values are not constant and Nakai \& Kuno (\cite{nakai}) 
find the following tendency: compared to our Galaxy, the $X$ values are smaller 
or comparable in galaxies earlier than Scd, and greater towards very late-type 
or irregular galaxies.  They also find that $X$ depends on $I_{\rm CO}$ as $X$ 
is larger for small $I_{\rm CO}$.  Metallicity, in addition, is a fundamental 
parameter determining the $X$ ratio, and is decreasing exponentially with 
radius in the Galaxy (Sakamoto \cite{sakamoto}).  From a study of 5 local group 
galaxies Wilson (\cite{wilson}) finds that $X$ significantly increases as the 
metallicity of the host galaxy decreases.

Our interferometric observations allow us to isolate molecular complexes in 
\object{M 81}.  Though these structures are bigger than typical Galactic GMCs, 
we try to derive a value of $X$ for these molecular complexes from two 
different methods: the virial theorem and the visual extinction $A_{\rm V}$.  
The latter method requires an assumption about the relative distribution of 
stars and gas along line-of-sight.

\subsection{Virial mass}\label{virial}

In order to calculate the virial mass, we used $M_{\rm vir}=95\,D_{\rm
e}\,\Delta V^{2}$ where $M_{\rm vir}$ is in M$_\odot$, $D_{\rm e}$ is the
effective diameter in pc and $\Delta V$ is in km\,s$^{-1}$ (Solomon et al.
\cite{solomon}).  This formula is obtained assuming that the cloud is a
sphere with a power-law ($\alpha $=1) density distribution.  If instead we
represent the cloud as a sphere with uniform density, the virial mass is then:
$M_{\rm vir}=104\,D_{\rm e}\,\Delta V^2$ and the mass is just 10\% higher.

To calculate the masses, we derived from the spectra the rms noise $\sigma $ 
and the linewidths at half intensity $\Delta V$.  And from the line contour 
maps we measured the size of the complex at a 3$\sigma $ level as well as the 
integrated intensity $I_{\rm CO}$ averaged over the cloud's area; an effective 
diameter $D_{\rm e}$ is then derived from the area of the complex (area=$\pi 
{{D_{\rm e}^2}\over{4}}$).  We make the hypothesis that the clouds are 
spherical so that there is no effect due to the inclination of the galaxy. The 
cloud diameters were deconvolved from the synthesized beam using a gaussian 
model.  Anyway as they are about twice the FWHM of the synthesized beam, the 
deconvolution just leads to a 10\% smaller virial mass and a 10\% smaller
value of $X$.

Because the clouds in this paper have a rather large sky-plane radius ($\sim 
100$ pc), it is likely that they are somewhat flattened.  The thickness of the 
molecular disk is 120 pc (FWHM) in the Milky Way (Combes \cite{combes}) and 
probably smaller in \object{M 81} because of the lower velocity dispersion.  
Nevertheless, the following example suggests that the virial mass of the cloud 
calculated by assuming these clouds are spherical is close to that obtained for 
a somewhat flattened cloud because the inclination correction tends to 
compensate for the oblateness correction.  The virial mass for an oblate 
spheroid of uniform density and eccentricity $e$ is the mass for the uniform 
density sphere multiplied by the factor $e$/arcsin $e$ (Binney \& Tremaine 
\cite{binney}).  The inclination correction to reduce the effective diameter to 
face-on (inclination $i = 0\degr$) increases from a factor of 1 for a sphere to 
a factor of (cos $i$)$^{-1/2}$ for a highly flattened spheroid.  The 
inclination $i$ is $59\degr$ in \object{M 81}.  As an example, if we suppose 
the clouds have a minor/major axis ratio in the range 0.2--0.6, with the minor 
axis perpendicular to the plane of \object{M 81}, then, to get the virial mass 
of a uniform, oblate, spheroidal cloud in \object{M 81} from the virial mass of 
a uniform sphere, one multiplies by a total correction factor (for oblateness 
and inclination) of 0.97--1.01, respectively.

 \begin{table*}
 \caption[]{Identified molecular complexes noted C1 to C6 (see
 fig.~\ref{integrated-map}).  The central coordinates are listed in Columns 2
 and 3.  Columns 4, 5, 7 give, respectively, the rms noise, the central
 velocity, and the linewidth at half intensity calculated from the spectra
 averaged over the area of the clouds and with a velocity resolution of
 2.6\,km\,s$^{-1}$.  Column 6 lists the central HI velocity (Adler, private
 communication).  Column 8 gives the effective radius in arcseconds deduced
 from the area of the complex (area=$\pi R_{\rm e}^2$) and deconvolved from
 the beam
 (1$^{\prime\prime}$ corresponds to 17.45\,pc).  Column 9 gives the peak
 brightness temperature measured from the spectra with a 2.6\,km\,s$^{-1}$
 velocity resolution (1\,Jy/beam = 3.8\,K for the 5$\,.\!\!^{\prime\prime}$1
 $\times$ 4$\,.\!\!^{\prime\prime}$8 synthesized beam).  Column 10 lists
the CO
 integrated intensity averaged over the clouds, column 11 the
 virial mass, and column 12 the deduced $X$ ratio.}

 \centering
\begin{tabular}{lr@{.}lr@{.}lrccr@{.}lr@{.}lr@{.}lcr@{.}lc}
 \hline
 \multicolumn{1}{c}{Cloud} &
 \multicolumn{2}{c}{$\alpha_{1950}$} &
 \multicolumn{2}{c}{$\delta_{1950}$} &
 \multicolumn{1}{c}{$\sigma$} &
 \multicolumn{1}{c}{$v_{\rm lsr}(\rm CO)$} &
 \multicolumn{1}{c}{$v_{\rm lsr}(\rm HI)$} &
 \multicolumn{2}{c}{$\Delta v$ } &
 \multicolumn{2}{c}{$R_{\rm e}$} &
 \multicolumn{2}{c}{$T_{\rm B}$} &
 \multicolumn{1}{c}{$I_{\rm CO}$} &
 \multicolumn{2}{c}{$M_{\rm vir}$} &
 \multicolumn{1}{c}{$N(\rm H_2)/I_{\rm CO}$} \\
 \multicolumn{1}{c}{} &
 \multicolumn{2}{c}{$^{\rm h}\,^{\rm m}\,^{\rm s}$} &
 \multicolumn{2}{c}{$\degr\,^{\prime}\,^{\prime\prime}$} &
 \multicolumn{1}{c}{mK} &
 \multicolumn{1}{c}{km\,s$^{-1}$} &
 \multicolumn{1}{c}{km\,s$^{-1}$} &
 \multicolumn{2}{c}{km\,s$^{-1}$} &
 \multicolumn{2}{c}{$^{\prime\prime}$} &
 \multicolumn{2}{c}{K} &
 \multicolumn{1}{c}{K\,km\,s$^{-1}$} &
 \multicolumn{2}{c}{M$_\odot$} &
 \multicolumn{1}{c}{$^{**}$} \\
 \multicolumn{1}{c}{(1)} &
 \multicolumn{2}{c}{(2)} &
 \multicolumn{2}{c}{(3)} &
 \multicolumn{1}{c}{(4)} &
 \multicolumn{1}{c}{(5)} &
 \multicolumn{1}{c}{(6)} &
 \multicolumn{2}{c}{(7)} &
 \multicolumn{2}{c}{(8)} &
 \multicolumn{2}{c}{(9)} &
 \multicolumn{1}{c}{(10)} &
 \multicolumn{2}{c}{(11)} &
 \multicolumn{1}{c}{(12)} \\
 \hline
C1  & 09 50 53&6 &  69 22 59&5 &  40  & 173.9 & 172.4 & 11&0 & 6&0  &
0&8 &
4.6 & 2&4 $10^6$ & 7.0 $10^{20}$ \\
C2  & 09 50 55&3 &  69 23 09&0 &  65  & 173.1 & 174.4 & 7&7 & 4&55  & 1&0 &
5.3 & 0&9 $10^6$ & 3.9 $10^{20}$ \\
C3$^*$  & 09 50 48&9 &  69 23 02&0 &  110 & 170.9 & 171.4 & 7&2 & 5&4 
& 1&3
& 5.7 & 0&9 $10^6$ & 2.7 $10^{20}$ \\
C4  & 09 50 53&4 &  69 22 38&7 &  70  & 175.7 & 175.0 & 7&2 & 4&5  &
0&4 & 2.5
& 0&8 $10^6$ & 7.3 $10^{20}$\\
C5$^*$  & 09 50 58&4 &  69 23 04&6 &  110 & 173.1 & 173.7 & 13&8 & 4&6 
& 0&5
&6.2 & 2&9 $10^6$ & 10.6 $10^{20}$\\
C6  & 09 50 55&6 &  69 22 44&2 &  55  & 168.2 & 168.8 & 6&9 & 5&25  & 0&3 &
1.6 & 0&8 $10^6$ & 9.0 $10^{20}$\\
 \hline
 \end{tabular}
  \begin{tabular}{l}
$^*$ Clouds C3 and C5 lie beyond the half power point of the primary beam and
the derived parameters are more uncertain \\
$^{**}$ in H$_2$\,cm$^{-2}$/(K\,km\,s$^{-1}$)\\
 \end{tabular}
 \label{table}
\end{table*}

If we assume that the virial mass represents the molecular hydrogen and helium 
mass, we can derive the ratio $X=N\,({\rm H}_2)/I_{\rm CO}$ and the results are 
listed in column 12 of Table 1.  The total uncertainty on these values is about 
50\%.

The mean value of $X$ is $6.7\pm 2.3\,10^{20}$ H$_2$cm$^{-2}$/(K\,km\,s$^{-1})$ 
for the six clouds.  The uncertainties are much bigger for complexes C3 and C5 
which are located on the border of the primary beam, and for complex C6 which 
has fainter emission.  Omitting these clouds with less well determined 
parameters, the mean value of $X$ is $6.1 \,10^{20}$ 
H$_2$cm$^{-2}$/(K\,km\,s$^{-1}$).  The mean value is about 3 times the standard 
Galactic value and about twice the value expected from the metallicity relation 
of Wilson (\cite{wilson}).  The metallicity of this field is close to the one 
measured in the solar neighborhood.  Indeed Garnett \& Shields (\cite{garnett}) 
find 12+log(O/H)=8.94 towards HII region 138 and a mean value of 8.79 at the 
galactocentric distance of this field in \object{M 81}, whereas Shaver et al.  
(\cite{shaver}) get 8.70$\pm $0.04 for the solar neighborhood.  For such 
metallicities, as for \object{M 31} and inner clouds in \object{M 33}, Wilson 
(\cite{wilson}) find $X = 3.3\,10^{20}$ H$_2$cm$^{-2}$/(K\,km\,s$^{-1})$.

Except for cloud C3, the value of $X$ is unusually large.  If the molecular 
hydrogen + helium mass represents the virial mass, such a large ratio means 
that, though CO emission in this \object{M 81} field is weak, there is a lot of 
H$_2$.  As the metallicity is normal, this could be due to a recent massive HI 
to H$_2$ conversion due to collisions; clouds C1, C2 and C5 are along the dust 
lane a little downstream from the HI ridge and may form a cloud collision front 
(Elmegreen \cite{elmegreen}).  We may be seeing only the warm envelopes of cold 
molecular clumps, so the integrated CO intensity may not represent the entire 
molecular mass of the cloud.  Alternatively, if the true $X$ value in \object{M 
81} is in fact comparable to the Galactic one, the virial mass is about 3 times 
bigger than the molecular mass and either the clouds are virialized because of 
a large mass of stars and/or cold atomic gas not taken into account in these 
calculations (see next paragraphs), or the clouds are not virialized.  In 
particular, Taylor \& Wilson (\cite{taylor}) recently made interferometric 
observations with OVRO of another field on the spiral arms of \object{M 81} 
with a higher angular resolution (3$''$).  They have identified three GMCs with 
diameters $\sim $100\,pc and very narrow linewidths of 4.7--7.7 km\,s$^{-1}$.  
Their field also has nearly solar metallicity, and their values for the 
integrated CO flux and $M_{\rm vir}$ imply that $X$ is $1.4 \pm 0.7\,10^{20}$ 
H$_2$cm$^{-2}$/(K\,km\,s$^{-1}$) in their clouds.  The GMCs' peak brightness 
temperatures are 3\,K and with beam dilution it is consistent with the mean 
1\,K peak brightness temperature of the complexes of this paper.  Thus it is 
possible that the molecular complexes identified in this paper contain several 
(2 or 3) gravitationally bound clouds of mass a few 10$^5$\,M$_\odot$ and size 
$\le $100\,pc not resolved in our data.  In the case of cloud C3, which is 
associated to the giant HII region (see sect.~\ref{HII}), we find the standard 
$X$ value and, though the derived parameters are more uncertain, it could be a 
single gravitationally bound cloud.

In the virial mass we have neglected the mass of the stars.  Indeed if we 
assume that the mass of the optical disk is about $10^{11}$\,M$_\odot$ within a 
radius of 20\,kpc, with an exponential distribution for the surface density 
$\mu = \mu_{0}exp(-r/r_{0})$, $r_{0}$=3\,kpc, the surface density 6.5\,kpc far 
from the center is $\mu = 200$\,M$_\odot$\,pc$^{-2}$.  This is comparable to a 
GMCs' typical mean surface density (170\,M$_\odot$\,pc$^{-2}$, Solomon et al.  
\cite{solomon}), but the stellar distribution has a scale height of about 10 
times that of the GMC, so that the volume density is about 10 times smaller for 
the stars than for the GMCs and we can neglect the mass of the stars at such a 
distance from the center.  However the molecular surface densities derived for 
the \object{M 81} clouds are smaller (30--150\,M$_\odot$\,pc$^{-2}$) with the 
derived $X$ ratio and even smaller (7--30\,M$_\odot$\,pc$^{-2}$) with the 
standard $X$ value.  Thus the mass of the stars may be less negligible compared 
to the molecular mass and may contribute to the binding of the clouds.

\begin{figure}[ht]
\includegraphics{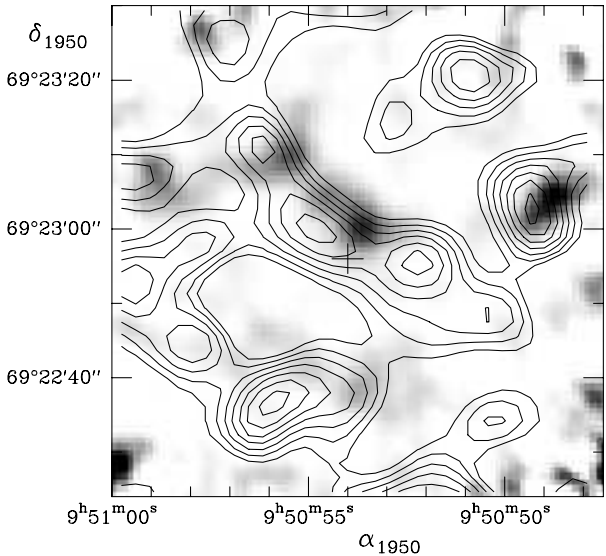}
\caption[]{HI contours (from Hine's map) superimposed on the
CO line integrated-intensity map (grey scale).  The first HI contour
corresponds to a column density of 1.9\,10$^{21}$\,atom\,cm$^{-2}$
(15\,M$_\odot$\,pc$^{-2}$) and the spacing is
0.3\,10$^{21}$\,atom\,cm$^{-2}$.
The resolution of the HI map is $9^{\prime\prime}$. See fig. 1 for the CO map
parameters.}
\label{HI/CO}
\end{figure}

In these calculations we have also neglected the HI gas.  However there is a 
good agreement between the CO and HI distribution (see fig.~\ref{HI/CO}) and 
velocities (see Table 1, where the HI velocities were measured by Adler on the 
HI image with $12''$ resolution from Adler and Westpfahl \cite{adler}).  It 
seems that there is a shift between the HI and CO maxima but the difference of 
resolution ($9^{\prime\prime}$ for the HI map of Hine (\cite{hine}) and 
$5^{\prime\prime}$ for the CO map) does not allow us to conclude about whether 
the shift is significant.  In section 4.2 we conclude that for cloud C1 the HI 
is not associated with the molecular cloud as it is mainly in a cold dust lane 
at high $z$; this is also true for cloud C2 since the overlaid HI is along the 
same HI feature.  In the case of cloud C3, it is clear that one should not 
include the entire HI column density in the cloud mass, because the resulting 
$M$(HI) alone times the 1.36 correction factor for helium would equal 
$M_\mathrm{vir}$ (see sect.~\ref{HII}).  For the other complexes, if we take 
into account the HI correction in the virial mass calculation and we assume 
that all the HI column density measured towards the CO complexes (Hine 
\cite{hine}) also contributes to the virial mass 
$M_\mathrm{vir}=M(\mathrm{H_2})+M(\mathrm{He})+M(\mathrm{HI})$, then we deduce 
a mean $X$ ratio of $4.3\pm1.8\,10^{20}$ H$_2$cm$^{-2}$/(K\,km\,s$^{-1})$, or 
omitting complexes C3, C5, and C6, which have more uncertain parameters, the 
mean $X$ is $3.0\,10^{20}$ H$_2$cm$^{-2}$/(K\,km\,s$^{-1})$.

Nevertheless, the most plausible explanation for the high values of X is that 
the complexes are not virialized but consist of a few unresolved
self-gravitating clouds.

\subsection{$A_v$/$N_{\rm gas}$}\label{av}

Kaufman et al.  (\cite{kaufmanb}) measured the extinction of several dust 
filaments in \object{M 81} and compared them with the HI surface density. One 
of their positions located at $\alpha_{1950} = 9^{\rm h} 50^{\rm m} 
53\,.\!\!^{\rm s}8, \delta_{1950} = 69\degr 23^{\prime} 
00\,.\!\!^{\prime\prime}3$ corresponds to our complex C1.  To analyze their 
data they used a dust-cloud model where the cloud is located at a height $z$ 
from the midplane and has a uniform distribution of dust.  For C1 they found 
$z$=200--400\,pc and $\tau_{\rm v}$=4--8.  The HI column density $N({\rm HI})$ 
is calculated from the HI map of Hine (\cite{hine}) averaged over the VLA 
9$^{\prime\prime}$ beam and $N({\rm HI})=(2.25\pm0.75) 
10^{21}$\,atom\,cm$^{-2}$ for C1.  If we know the $A_{\rm v}/N_{\rm gas}$ 
ratio, we can deduce $N_{\rm gas}$ and then derive the molecular amount $N({\rm 
H_{2}})$ necessary to account for the extinction.

The value of $N_{\rm gas}/E(B-V)$ varies from one galaxy to another because of 
the metallicity (Nakai \& Kuno \cite{nakai}), but the value of $A_{\rm 
v}/E(B-V)$ should not vary with galaxies if the dust properties are the same on 
a large scale and in a moderately dense medium.  So we can use the Galactic gas 
to extinction ratio (Bohlin et al.  \cite{bohlin}) corrected for metallicity 
(as in Kaufman et al.  \cite{kaufmana}).  We thus have $A_{\rm v}/N_{\rm 
gas}$=0.58\,$10^{-21}$\,mag\,atom$^{-1}$\,cm$^2$ at $R$=6\,kpc.

With $N_{\rm gas}=1.9\,10^{21}\,\tau_{\rm v}$\,atom\,cm$^{-2}$, then using the 
HI column density measured with $9''$ resolution and assuming all the gas is in 
front, we have $N({\rm H_2})$\,=\,$N_{\rm gas}$\,--\,$N({\rm 
HI})$=5.2\,10$^{21}$atom\,cm$^{-2}$ for $\tau_{\rm v}$=4 or 
1.3\,10$^{22}$atom\,cm$^{-2}$ for $\tau_{\rm v}$=8.  From our data we measured 
a CO integrated intensity $I_{\rm CO}$=4\,K\,km\,s$^{-1}$.  For this extinction 
comparison, we give the integrated CO intensity centered on the dust position 
and averaged over the 9" FWHM of the HI beam, whereas Table 1 lists the 
integrated intensity averaged over the cloud.  If we assume that the 
interferometer missed at most 30\% of the CO flux, we get $X$ in the range
4.5\,10$^{20}$ to 15\,10$^{20}$\,H$_2$cm$^{-2}$/(K\,km\,s$^{-1}$).  If we 
average the CO emission over just the width of the dust lane, then $I_{\rm CO}$ 
= 7\,K\,km\,s$^{-1}$, and assuming that the interferometer detected all the 
emission in this small region, we get $X$ = 
3.5--8.8\,10$^{20}$\,H$_2$cm$^{-2}$/(K\,km\,s$^{-1}$).  These various estimates 
are consistent with the values of $X$ derived above.

We assumed here that the molecular cloud is associated with the dust lane and 
thus at a height $z$=200--400\,pc, which is unlikely.  The molecular scale 
height in our Galaxy is about 50\,pc and the star formation is more violent 
than in \object{M 81}, so that the stellar winds are more present to push 
clouds up.

An extinction of $\tau_{\rm v}$=4 implies a gas column density of 
7.6\,10$^{21}$\,atom\,cm$^{-2}$.  If the HI were concentrated in the linear 
dust lane of width 2.5$^{\prime\prime}$, the HI column density would be 
8.1\,10$^{21}$\,atom\,cm$^{-2}$, enough to explain alone the extinction.  But 
the HI column density is similar towards neighboring regions, so there is no 
evidence for the HI to be confined in the dust lane.  For cold and dense HI 
clouds with $N({\rm HI})$=8\,10$^{21}$\,atom\,cm$^{-2}$, a typical $\Delta 
v$=10\,km\,s$^{-1}$ and temperature $T$=10\,K, the HI is highly optically thick 
and we don't measure the true HI column density.  We can finally assume that 
there is both diffuse optically thin HI (at a typical temperature of 100\,K) 
and the HI confined to the dust lane at a high $z$ and optically thick (at a 
typical temperature of 10\,K).  It can be the same for the CO emission as we 
detect the less optically thick emission and we don't see the gas providing 
most of the extinction.

We can thus account for the observed extinction given the measured HI and CO
emissions. However, we note that the dust position in cloud C1 is not one of
the positions measured by Kaufman et al. (\cite{kaufmanb}) with a large
discrepancy between the extinction and $N({\rm HI})$.

\section{Cloud C3 and HII Region 138} \label{HII}

Cloud C3 and the Taylor \& Wilson (\cite{taylor}) cloud MC2 represent the first 
identifications in \object{M 81} of CO emission associated with an individual 
giant HII region (see fig.  1).  Previous CO observations of \object{M 81} had 
too low a spatial resolution to connect the weak CO emission with the 
star-forming regions, so it has long been a puzzle as to how \object{M 81} is 
forming massive OB associations in the presence of so little molecular gas.  
The answer for the interferometer field of fig.  1 is that most of the 
molecular mass, as indexed by the CO emission, is in the form of 
$10^5-10^6$\,M$_\odot$ complexes, suitable as birthplaces for large OB 
associations.

The giant HII region 138, with excitation parameter U\,=\,250\,pc\,cm$^{-2}$ 
and diameters 17$^{\prime\prime}$\,x\,13$^{\prime\prime}$ (Kaufman et al.  
\cite{kaufman}) appears inside cloud C3.  In projection, the center of the 
cloud complex is displaced 7$^{\prime\prime}$\,=\,120\,pc from the center of 
the HII region.  The location of the HII region is suggestive of a blister; 
possibly, the molecular gas on the eastern side of the HII region was 
photodissociated by the OB stars.  Also, Kaufman et al.  (\cite{kaufman}) 
conjecture that a nonthermal radio source adjacent to the southeastern side of 
the HII region may be an SNR, which would have affected the molecular cloud.  
In the Milky Way, individual HII regions with free-free radio luminosity 
similar to that of HII region 138 are associated with molecular clouds of mass 
4\,10$^5$ to 4\,10$^6$\,M$_\odot$ (Myers et al.  \cite{myers}).  If cloud C3, 
associated with HII region 138, is a single virialized cloud, its mass 
(0.9\,10$^6$\,M$_\odot$) lies within this range.  HII region 172 (Kaufman et 
al.  \cite{kaufman}) near cloud MC2 in the Taylor \& Wilson (\cite{taylor}) 
field has a free-free radio luminosity similar to HII region 138 , but MC2 has 
a virial mass of only 3\,10$^5$\,M$_\odot$, i.e., at the low end of the Milky 
Way range for clouds related to such luminous HII regions.  Integrating the 
entire HI column density over the solid angle of the HII region gives an HI 
mass $M$(HI) of 1.1\,10$^6$\,M$_\odot$ for HII region 138 and 
2.3\,10$^6$\,M$_\odot$ for HII region 172 (Kaufman et al.  \cite{kaufmana}), 
not corrected for helium.  In both cases, the molecular cloud is displaced 
towards the western side of the HII region, and the atomic mass in the region 
exceeds the molecular mass.

A layer of gas with column density 1.2\,10$^{21}$\,atom\,cm$^{-2}$ in front of 
the HII region would suffice to explain its observed extinction, $A_{\rm 
v}$\,=\,0.7$\pm$0.3\,mag.  Averaged over the HII region, the CO integrated 
intensity is 1.9\,K\,km\,s$^{-1}$, which corresponds to a total column density 
through the disk of 5\,10$^{20}$ H$_2$\,cm$^{-2}$ if we use the value of $X$ = 
2.7\,10$^{20}$ H$_2$cm$^{-2}$/(K\,km\,s$^{-1}$) found for cloud C3, and the 
HI column density through the disk is 2.4\,10$^{21}$\,atom\,cm$^{-2}$.  Thus 
there is more than enough gas to account for the observed extinction, and the 
HII region is probably either at the midplane of the gas layer or slightly to 
the near side.

\section{Conclusion}

We have mapped and identified in \object{M 81} six molecular complexes with 
virial masses $\sim$ 1--3\,10$^6$\,M$_\odot$ and radii $\sim 100$\,pc in an 
optimally-chosen field of deprojected size 1.1 $\times$ 2.1\,kpc (the 
inclination of \object{M 81} is 59$\degr $).  Comparing the virial masses with 
the CO fluxes of these clouds, we find a mean value of the $N\,({\rm 
H}_2)$/$I_{\rm CO}$ ratio 3 times stronger than the standard Galactic value and 
2 times more than the value expected for such metallicities.  The most 
plausible explanation is that the complexes are the average of several 
gravitationally bound clouds not resolved in our data.  A possible exception is 
cloud C3 which is associated to the giant HII region and, though the derived 
parameters are more uncertain, it could be a single gravitationally bound 
cloud.  Severe beam dilution in the single-dish surveys (e.g., with the NRAO 
12\,m telescope) appears to account, in part, for the weakness of the overall 
CO emission from \object{M 81} in such surveys.

However we observed here an optimal field as it overlapped one of the stronger 
fields in the CO single-dish surveys.  Its CO emission may not be typical of 
\object{M 81} as a whole: the sum of the virial masses of the six complexes is 
8.7\,10$^6$\,M$_\odot$, which gives an upper limit to the H$_2$ mass of 
6.4\,10$^6$\,M$_\odot$ if helium is omitted.  (On the other hand, if the 
standard value of $X$ applies, the H$_2$ mass is 2.3\,10$^6$\,M$_\odot$, 
omitting helium.)  The upper limit is $\sim$6\% of the
total molecular mass of \object{M 81}.  The corresponding HI mass of this field is 
1.65\,10$^7$\,M$_\odot$, which leads to a molecular to atomic gas ratio of 0.14 
if we use the standard value of $X$ and neglect a diffuse component, or up to 
0.6 if we use the sum of the virial masses and take the upper limit on the 
diffuse component from the comparison of the 30\,m and interferometer data.  
Furthermore this field contains a giant radio HII region coinciding with one of 
the molecular complexes and the other clouds are on the UV-bright arm, so that 
there is adequate heating of the gas.  But the observed narrow linewidths have 
still to be explained.

The upper limit to the average molecular column density in this optimum field 
on the arm (obtained from the sum of the Virial masses) is approximately 
6\,M$_\odot$\,pc$^{-2}$.  This is much smaller than even the interarm surface 
density of molecular gas in the CO-rich galaxy \object{M 51} (see Rand et al. 
\cite{rand}).

The 3 GMCs in M81 detected by Taylor and Wilson (\cite{taylor}) in another 
field of the spiral arms have CO integrated fluxes of 1.6--3.2 
Jy\,km\,s$^{-1}$, diameters of $\sim$ 100\,pc, and linewidths of 
5--8\,km\,s$^{-1}$.  For comparison, our 6 complexes have CO integrated flux of 
1.3--5.0 Jy\,km\,s$^{-1}$, diameters of 160--200 pc, and linewidths of 
7--14\,km\,s$^{-1}$.  Thus our complexes have somewhat similar 
$^{12}$CO\,(1--0) fluxes but significantly larger areas.  In both cases, the 
linewidths are narrow as compared to Galactic clouds of the same diameter.  
This has some interesting implications.  Because of the low velocity 
dispersion, it is possible that the scale height of the molecular gas is 
smaller than in the Milky Way.  However, if the complexes are not significantly 
flattened, then the mean volume density in the M81 clouds must be less than in 
Galactic clouds.  This would mean that the clouds themselves are different from 
those in the Milky Way.  Taylor and Wilson (\cite{taylor}) speculate that their 
clouds might be complexes consisting of several subclouds, like the Orion 
complex.  This is one way in which the mean volume density could be reduced.

 From our interferometric and single-dish data, it seems that this field in 
\object{M 81} has no fewer molecular complexes than other comparable galaxies 
but lacks a significant widespread distribution of small molecular clouds (in 
the limits of the calibration uncertainties, i.e.  less than about 30\% of the
total flux).  This lack of small molecular clouds could be due, in this 
particular field, to a collision front (see sect.~\ref{virial}) where the small 
clouds have agglomerated into big ones.  Another possibility is that, as the 
molecular surface density in \object{M 81} is low, the small 
diffuse clouds have less self-shielding against CO photodissociation than 
the GMCs.

In the regions of \object{M 81} with little UV emission (e.g. the interarm 
regions and parts of the central 4\,kpc HI hole), there is little heating of 
the gas, so most of the CO gas there may be too cold to have been detected in 
the surveys thus far.  Allen et al.  (\cite{allen2}) reach the same conclusion 
from their picture of HI as a product of star formation process, and they 
speculate that the gas in the main part of the disk of \object{M 81} is largely 
in the form of molecular hydrogen with a low excitation temperature.  It is 
still uncertain whether the $^{12}$CO\,(1--0) emission from \object{M 81} as a 
whole can be used directly as a reliable tracer of the molecular gas.

We conclude from the observations presented here that the molecular medium 
in \object{M 81} differs from that in the Milky Way.\\

\begin{acknowledgements}

We thank the IRAM staff in Grenoble and Granada for their help. We thank too
David Adler for measuring the HI velocities.  We acknowledge our referee
Christine Wilson for her very helpful comments.

\end{acknowledgements}

\end{document}